\documentclass{epl}

\begin{document}

\title{Dissipate locally, couple globally: a sharp transition from
decoupling to infinite range coupling in Josephson arrays with
on-site dissipation} \shorttitle{Floating to coupled transition in
Josephson arrays with on-site dissipation}
\author{ S. Tewari\inst{1} \and J. Toner\inst{2}}

\institute{
   \inst{1} Condensed Matter Theory Center, Department of Physics,
   University of Maryland, College Park, MD 20742\\
   \inst{2} Department of Physics and Institute of Theoretical
Science, University
                   of Oregon, Eugene, OR 97403
} \pacs{74.81.Fa}{Josephson junction arrays and wire networks}
\pacs{71.10.Hf}{Non-Fermi-Liquid ground states}
\pacs{74.78.Na}{Mesoscopic and nanoscale systems}

\maketitle

\begin{abstract}
We study the $T=0$ normal to superconducting transition of
Josephson arrays with {\it on-site} dissipation. A perturbative
renormalization group solution is given. Like the previously
studied case of {\it bond} dissipation (BD), this is a ``floating"
to coupled (FC) phase transition. {\it Unlike} the BD transition,
at which {\it only} nearest-neighbor couplings become relevant,
here {\it all} inter-grain couplings, out to {\it infinitely}
large distances, do so simultaneously. We predict, for the first
time in an FC transition, a diverging spatial correlation length.
Our results show the robustness of floating phases in dissipative
quantum systems.
\end{abstract}

Coupling dissipation to quantum systems has interesting
consequences. Quite generally, dissipation suppresses quantum
fluctuations. This is evident at the level of a single macroscopic
quantum `particle' whose tunnelling probability out of a
metastable state is suppressed by dissipation \cite{Caldeira}. For
the  quantum particle in a double well \cite{Sudip1, Bray}, or in
a periodic \cite{Schmid, Fisher1} potential, Ohmic (i.e., linear)
dissipation can suppress quantum fluctuations enough to cause a
dissipative `quantum to classical' phase transition. In the
classical phase, all quantum fluctuations are quenched and the
system is spontaneously trapped into only one of the potential
minima.

   From the point of view of understanding effects of dissipation on
extended systems, a natural question to ask is what happens when
such zero-dimensional dissipative systems are spatially coupled in
a lattice. Understanding such effects has important practical
applications. Such dissipative effects are thought to be at the
heart of the physics of granular superconductivity \cite{CIKZ}.
They are also important in the context of decoherence in a qubit,
which, it has been proposed\cite{Ioffe}, can be realized in a
Josephson junction array . If the coupling of dissipation remains
local,
then it is natural to expect that quantum fluctuations can also be
{\it locally} quenched by dissipation. At zero temperature,
coupling such classical systems spatially should give rise to an
ordered state of the extended system. This dissipative quantum
phase transition (QPT), however, is expected to retain some of its
local character.

Early work on these questions\cite{CIKZ, Fisherdiss, Zwerger1,
Panyu, Zwerger2, Zant, Ising} was recently extended in
Ref.~\cite{Sumanta1} in the context of a Josephson array with
dissipation coupled to the {\it bond} phases, that is, the order
parameter phase {\it differences} of two consecutive
superconducting grains; see also Ref.~\cite{Gill}. This situation
can be realized when the junctions are shunted by Ohmic resistors.
It was shown that the metal to superconductor dissipation-tuned
QPT occurs via a floating to coupled phase transition. Interesting
local behavior in the bond dissipation model has also been
addressed in Ref.~\cite{Bobbert}.

The case where dissipation couples to the site variables - that
is, the phases of the superconducting islands - directly has also
been addressed in the literature \cite{Wagenblast, Klaus, Polak}.
Such local damping of the phase happens \cite{Wagenblast} when
Cooper pairs  can decay into localized pools of normal electrons
in the substrate. It can also be built into an artificial array
with each island shunted separately to the ground \cite{Polak}.
Such a `shunt to the ground'  also occurs naturally in many
granular systems. The model has been mapped on to effective LZW
field theories \cite{Wagenblast, Polak} to study the normal to
superconductor phase transition.

\begin{figure}[htb]
\centerline{\includegraphics[scale=0.7]{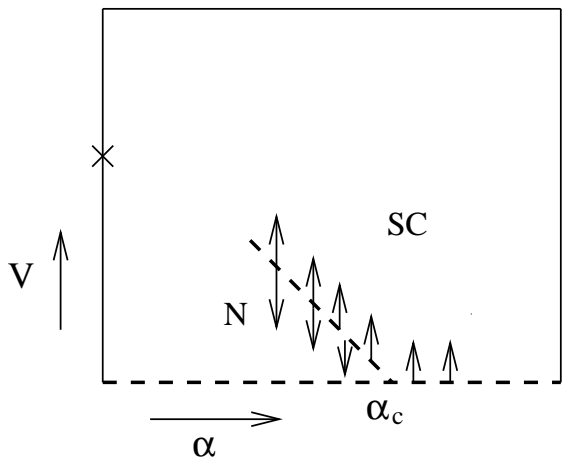}}
\caption{Phase diagram and perturbative RG flows in the $V-\alpha$
plane in any dimension $D>0$. N and SC indicate normal and
superconducting phases, respectively. The normal metallic phase is
the floating phase. The thick dashed lines correspond to lines of
fixed points. The cross indicates a special point on the
$\alpha=0$ axis denoting a ($D+1$)-dimensional $XY$ transition.
The part of the $\alpha=0$ axis from the cross to infinity is a
line of fixed points in $D=1$.}\label{Fig1}
\end{figure}

Working directly with the order parameter {\textit {phase}}
variables, we give in this paper a second order perturbative
renormalization group (RG) solution for the site dissipation
problem. We show that a floating normal metallic phase, which is
even more striking than in the bond-dissipation case, and a
floating to coupled QPT occur in this case. In contrast to
bond-dissipation, here {\it all} couplings between grains become
relevant simultaneously, and the transition is controlled by a
fixed, infinite-dimensional critical hypersurface with
continuously varying critical exponents whose dependence on
dissipation we calculate near the threshold dissipation. We also
predict, for the first time in a floating to coupled phase
transition (three other examples of which are known \cite{Toner,
Steve,Sumanta1}), a diverging spatial correlation length. The RG
flows and phase diagram we derive are displayed in Fig.~\ref{Fig1}
in the $V-\alpha$ plane, where $V$ is the nearest neighbor
Josephson coupling and $\alpha$ is the coefficient of dissipation.

\vspace*{0.3in}

Our results demonstrate the model-independence of floating phases
in dissipative Josephson arrays. They also indicate that
dissipative quantum systems may be the best places to
experimentally look for such phases, which may otherwise be very
difficult to obtain \cite{Coleman}. Moreover, they provide a rare
example in statistical mechanics of a critical fixed locus that
exists in any dimension; there is no upper critical dimension in
the problem.

On-site dissipation coupled to XY spins has also been treated in
Refs.~\cite{Subir1, Subir2}. There, dissipation is coupled locally
to a $\phi^4$-theory, which retains the compactness between $0$
and $2\pi$ of the phases of the order parameter. The problem is
then studied using an $\epsilon$-expansion around the upper
critical dimension. In our model, the compactness is explicitly
broken, as is expected in Ohmic dissipation \cite{Schon}. We
believe the difference of our results from Refs.~\cite{Subir1,
Subir2}
is due to this different treatment of the phase variable.
















The action for the Josephson junction array coupled to on site
Ohmic dissipation reads
\begin{eqnarray}
{\cal{S}}/\hbar&=& \sum_{\vec{r}}\sum_{n}(\frac{C}{2}\omega_n^2+
\frac{\alpha}{2\pi}|\omega_{n}|)
      |\tilde{\theta}(\vec{r},\omega_{n})|^{2 }\nonumber\\&+&
V\sum_{\langle \vec{r},\vec{r}'\rangle}\int
d\tau[1-\cos\Delta\theta_{\vec{r},\vec{r}'}(\tau)].
\label{action1}
     \end{eqnarray}

Here $C$ is the capacitance of the grains arranged on an arbitrary
$D$-dimensional lattice with lattice points $\vec{r}$.
$\theta({\vec{r}},\tau)$ is the phase of the superconducting order
parameter on the grain at $\vec{r}$ at imaginary time $\tau$, and
$\tilde{\theta}(\vec{r},\omega_n)$ is its temporal Fourier
transform. The bosonic Matsubara frequency $\omega_n={2\pi
n/\beta}$ with $n$ an integer, and $\beta$ the
inverse-temperature. $\langle \vec{r},\vec{r}'\rangle$ denotes
nearest neighbor pairs, and $\Delta
\theta_{\vec{r},\vec{r}'}=\theta(\vec{r})-\theta(\vec{r}')$. The
term containing $\alpha$ arises from integrating out a
harmonic-oscillator bath with the linear spectral-function
required to produce Ohmic resistance\cite{Caldeira}. In
$\tau$-space, this term reads $\frac{\alpha}{2\pi}\int d\tau\int
d\tau'(\frac{\theta(\vec{r},\tau)-\theta(\vec{r},\tau')}{\tau-\tau'})^2$,
which is a long ranged interaction in time. We restrict the sum on
the Matsubara frequencies to those satisfying $|\omega_n| <
\omega_c$, where $\omega_c$ is an ultraviolet cutoff .

To study the dissipation-tuned quantum phase transition in this
model, we use a renormalization group that is perturbative in the
spatial coupling $V$ \cite{CIKZ, Sumanta1}. This RG procedure
starts by dividing the field $\theta(\vec{r},\tau)$ into slow and
fast components $\theta_{s}(\vec{r},\tau)$ and
$\theta_{f}(\vec{r},\tau)$, respectively,  defined by having the
temporal Fourier transform of $\theta_{f}(\vec{r},\tau)$ equal
that of $\theta (\vec{r},\tau)$ for ``high'' Matsubara frequencies
obeying ${\omega_c\over b}<|\omega_n| < \omega_c$, and equaling
zero for all smaller Matsubara frequencies, while the temporal
Fourier transform of $\theta_{s}(\vec{r},\tau)$ equals that of
$\theta (\vec{r},\tau)$ for ``low'' Matsubara frequencies obeying
$|\omega_n| < {\omega_c\over b}$. We can then
%
%
%
%
%
%
write the partition function $Z$ as
     \begin{equation}
Z=Z_{0} \int\prod_{\vec{r}}{\cal
D}\theta_{s}(\vec{r},\tau)\exp{\left[-{\cal S}_{0}^s/\hbar +
\ln\langle e^{-{\cal S}^{\prime}/\hbar}\rangle_{0f}\right]}.
\label{partition}
     \end{equation}
Here $Z_{0}$ is a normalization constant, ${\cal S}_{0}^{s}$ is
the slow-frequency component of the quadratic part  of the action,
${\cal S}^{\prime}$ contains the spatial coupling term $V$, and
$\langle\ldots\rangle_{0f}$ denotes averages over the fast
components. After computing the averages, we rescale $\tau$,
$\tau'=\tau/b$, $b$ being the scale factor, to restore the
original frequency cut-off. Finally we redefine the coupling
constants to complete the renormalization. Note that we do {\it
not} rescale the fields $\theta({\vec{r}},\tau)$; this is to avoid
introducing a factor multiplying
$\Delta\theta_{\vec{r},\vec{r}'}(\tau)$ in the argument of the
cosine in Eq. (\ref{action1}). With this choice to not rescale
$\theta({\vec{r}},\tau)$, the dissipation term $\alpha$ in
Eq.~(\ref{action1}) is dimensionless. Furthermore, because it is
non-analytic in $\omega_n$, it gets no ``graphical'' corrections
(i.e., it is unchanged by the step of integrating out the ``fast''
fields). Hence, it is held fixed by the RG. The term associated
with $C$ has $\tau$-dimension $-1$; it is irrelevant in the RG
sense. The only place where $C$ appears is in the frequency
integrals such as,
\begin{equation}
\int_{|\omega|>\frac{1}{\beta}}^{\infty}\frac{d\omega}{2\pi}\frac{1}{C\omega^2
+\frac{\alpha}{2\pi}|\omega|}=\frac{1}{\alpha}\ln\frac{\alpha\beta}{\pi
C}. \nonumber
\end{equation}
Thus, $\frac{1}{C}$ acts as an upper frequency cut off, and we
associate it with $\omega_c$, $\omega_c=\frac{\alpha}{\pi C}$.

Writing  $b=e^l$, where $l>0$ is infinitesimal, we have, to one
loop order \cite{CIKZ, Sumanta1}:
\begin{equation}
\frac{dV}{dl}=(1-\frac{1}{\alpha})V \label{Vrg}
     \end{equation}
This rcursion relation implies that the $\alpha$-axis
($V=0$) is a line of stable fixed points for $\alpha<1$ and a line
of unstable fixed points for $\alpha>1$. Defining $\alpha_c=1$,
for $\alpha<\alpha_c$, the barrier $V$ between the different
potential minima of the cosine potential is irrelevant,
and so the fields $\theta(\vec{r},\tau)$ on each site fluctuate
quantum mechanically independently of those on all the other
sites. For $\alpha>\alpha_c$, on the other hand, the barrier $V$
grows, quenching these fluctuations, and the system becomes a
global superconductor.
Therefore, at $\alpha=1$, there is a metal-to-superconductor
quantum phase transition.

Before we go to the higher order corrections to Eq.~\ref{Vrg} and
discuss the phase diagram,
    we first show that on the metallic side of $\alpha=1$, the local
phases $\theta(\vec{r},\tau)$, which are critical in the imaginary
time variable, exist in a so called `floating' phase. The floating
phase for the array is defined as a phase where {\it all} spatial
Josephson couplings, including the longer ranged ones that
nominally couple spatially far separated grains, are irrelevant
\cite{Toner, Sumanta1}. From Eq.~\ref{action1} with irrelevant
$V$, the average of the order parameter on a grain is zero at
$T=0$, $\langle \exp(i\theta(\vec{r},\tau))\rangle = 1/(\omega_c
\beta)^{\frac{1}{2\alpha}}$. The on site unequal time correlation
function decays algebraically,
\begin{equation}
\langle \exp i(\theta(\vec{r},\tau)-\theta(\vec{r},0))\rangle =
\frac{1}{(\omega_c \tau)^{\frac{1}{\alpha}}}
\label{onsitecorrelator}.
\end{equation}
Now, a generic arbitrary range Josephson coupling among pairs of
grains takes the form:
\begin{equation}
{\cal S}_J = -\int d\tau \sum_{\vec{r},\vec{r}'}J(\vec{r}')
\cos(\theta(\vec{r},\tau)-\theta(\vec{r}+\vec{r}',\tau))\label{Sj},
\end{equation}
where $|\vec{r'}|$ gives the spatial range of the coupling. When
$V$ is irrelevant, the average of ${\cal S}_J$ scales as $\langle
{\cal S}_J \rangle = \beta \langle
\exp(i\theta(\vec{r},\tau))\rangle^2 \sim
\beta^{1-\frac{1}{\alpha}}$.  At zero temperature, this term is
irrelevant for $\alpha < 1$. Thus, all spatial couplings,
including the nearest neighbor coupling $V$, are irrelevant
throughout the metallic phase. Interestingly, they all become
relevant simultaneously at $\alpha=1$. Hence, for $\alpha<1$, the
system is in a floating critical phase.

The first order flow equation, Eq.~\ref{Vrg}, only establishes a
phase transition. To know the phase diagram and the critical
properties, such as correlation length and time exponents, we have
to go beyond first order. To obtain higher order corrections, we
perform a cumulant expansion of Eq.~\ref{partition} in second
order in $V$. By the method outlined in Ref.~\cite{Sumanta1}, we
find higher order contributions to Eq.~\ref{Vrg} coming from the
closed loops on the lattice. At higher orders in RG, longer ranged
pairwise interactions $J({\vec{r}})$ of Eq.~\ref{Sj} are
successively generated. In a consistent treatment of the RG, all
these terms should be included in the starting Hamiltonian. Since
the quadratic part of Eq.~\ref{action1} decouples in space, it is
easy to see that all the $J(\vec{r})$'s have the same linear order
recursion relation as that of the nearest neighbor interaction
$V$, Eq.~\ref{Vrg}. At second order cumulant expansion of
(\ref{partition}), $J(\vec{r})$ for a particular $\vec{r}$ is
generated by all possible combinations of $J(\vec{r}')$ and a
$J(\vec{r}-\vec{r}')$ for which $\vec{r}, \vec{r}'$ and
$\vec{r}-\vec{r}'$ lie on a triangle on the lattice
\cite{Sumanta1}. Treating $V$ as a special case of $J(\vec{r})$
(that corresponding to the nearest neighbor vectors $\vec{a_i}$,
$J(\vec{a_i})=V$), we get, up to quadratic order in the couplings,
\begin{equation}
\frac{dJ(\vec{r})}{dl}=(1-\frac{1}{\alpha})J(\vec{r}) + A
\sum_{\vec{r}'}J(\vec{r}')J(\vec{r}-\vec{r}').\label{Jrg}
\end{equation}
Here $A$ is a positive number, $A\sim
\int_0^{\beta}d(\tau-\tau')[\exp(-G^f(\tau-\tau'))-1]$, where
$G^f(\tau-\tau')=<\theta_f(0,\tau)\theta_f(0,\tau')>_{0f}$ is the
unequal time correlation function of the fast modes
\cite{Sumanta1}. To avoid generation of spurious long ranged
interactions, a smooth cut off prescription is adopted to
calculate $G^f (\tau-\tau')$, which then makes it exponentially
decaying with $|\tau-\tau'|$ \cite{Ma, Fisher1}. Since $A$ depends
on this cut off scheme, physical quantities such as exponents
should be independent of $A$.

It is essential for our subsequent analysis that all of the
factors $J(\vec{r}')J(\vec{r}-\vec{r}')$ enter the recursion
relation for $J(\vec{r})$ with the {\it same} coefficient $A$.
This is a simple consequence of the fact that, in contrast to the
{\it bond} dissipation case, here, in the absence of the Josephson
coupling terms, fluctuations of $\theta(\vec{r}, \tau)$ on any
site $ \vec{r}$ are completely independent of those on all other
sites $ \vec{r} '$.  As a result, the averages over
$\theta(\vec{r}, \tau)$'s on different sites that must be computed
in the perturbation theory to obtain $A$ are the same, regardless
of which sites $\vec{r}$ and $\vec{r}-\vec{r}'$ are being
considered.

Because of the fact that $A$ is a constant that factors out of the
sum on $\vec{r}'$ in (\ref{Jrg}), that formidable looking set of
infinitely many coupled non-linear recursion relations  decouple
in Fourier space. Defining the Fourier transform $J(\vec{q})$ of
$J(\vec{r})$ via $J(\vec{r})\equiv\int_{BZ}d^{d}q
J(\vec{q})\exp(i\vec{q}\cdot\vec{r})$ where the $\int_{BZ}d^{d}q$
is over the Brillouin zone of the lattice, and defining
$\epsilon\equiv{1\over\alpha} - 1$, Eq.~\ref{Jrg} reduces to
\begin{equation}
\frac{dJ(\vec{q})}{dl}=-\epsilon J(\vec{q}) + A J^2(\vec{q}).
\label{Jqrg}
\end{equation}

\begin{figure}[htb]
\centerline{\includegraphics[scale=0.5]{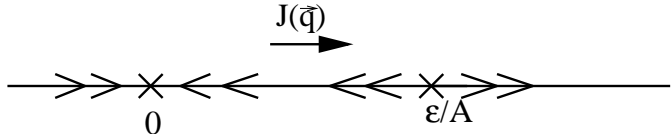}} \caption{RG
flows for $J(\vec{q})$ as implied by Eq.~\ref{Jqrg}.}\label{Fig2}
\end{figure}

Let's consider the metallic side of $\alpha_c=1$, where $\epsilon$
is positive. Equation~\ref{Jqrg} has a stable fixed point at
$J(\vec{q})=0$, and a critical fixed point at
$J(\vec{q})=\epsilon/A$. Fig.~\ref{Fig2} shows the flow of
$J(\vec{q})$.
\vspace*{0.5in}
\vspace*{0.5in}

\begin{figure}[htb]
\centerline{\includegraphics[scale=0.5]{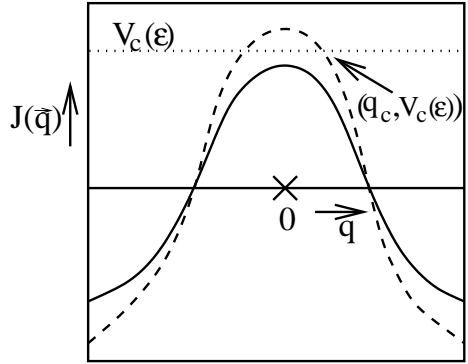}} \caption{A
typical form of the bare $J(q)$ in $q$-space. The dotted
horizontal line indicates $V_c(\epsilon)$. When the maximum of
$J(q)$ at $q=0$ lies below the dotted line (solid curve), the
system is in the floating metallic phase. When the maximum lies
above the dotted line (dashed curve), $J(q)$'s in a band of $q$
values, $|q|<q_c$, become relevant, and the system becomes a
global superconductor. }\label{Fig3}
\end{figure}

It follows that as long as the maximum of $J(\vec{q})$ in
$q$-space is less than $V_c(\epsilon)=\epsilon/A$, the system is
in the floating metallic phase. As shown in Fig.~\ref{Fig3}, when
the maximum lies above $V_c(\epsilon)$, $J(\vec{q})$'s for a band
of $q$ values become relevant and the system becomes a global
superconductor. To illustrate this, let's consider a particular
form of the bare $J(\vec{r})$, namely, that for nearest neighbor
interactions only, on a $D$-dimensional hypercubic lattice,
$J(\vec{r},l=0)=V\sum_{\vec{a_i}}\delta(\vec{r}-\vec{a_i})$, where
the $\vec{a_i}$'s are the set of nearest neighbor vectors on this
lattice. This gives a bare $J(\vec{q})$ of the form,
\begin{equation}
J(\vec{q},l=0)=V\sum_{\vec{a_i}}\cos(\vec{q}\cdot\vec{a_i}).\label{Jq}
\end{equation}

As shown in Fig.~\ref{Fig3}, as long as $zV$ is less than
$V_c(\epsilon)$, where $z=2d$ is the coordination number of the
lattice, all $J(\vec{q})$'s are irrelevant. When $zV$ is greater
than $V_c(\epsilon)$, $J(\vec{q})$'s in a band of $q$ values are
relevant. We expand $J(\vec{q},l=0)$ in the neighborhood of $q=0$,
\begin{equation}
J(\vec{q},l=0)\simeq J(q=0)-B|q|^2, \label{Jqexpand}
\end{equation}
where $J(q=0)=zV$  and $B=2V a^2$ (with $a$ the length of a
nearest neighbor vector) in the current example. For more general
lattice types, and further than nearest neighbor interactions, the
{\it form} of Eq.~\ref{Jqexpand} will still hold for small
$\vec{q}$, provided only that the bare Josephson couplings
$J(\vec{r})$ are such as to produce a net {\it ferromagnetic}
interaction (by which we mean one that favors a spatially uniform
$\theta(\vec{r}, \tau)$), and short-ranged. Indeed, that
$J(\vec{q})$ has its maximum at $\vec{q} = \vec{0}$ , and that
that maximum is a simple quadratic maximum as in
Eq.~\ref{Jqexpand}, can be taken as a {\it definition} of
ferromagnetic, short-ranged interactions. Furthermore, it will be
satisfied by any system that has, in the bare Hamiltonian, only
nearest-neighbor
%
%
%
ferromagnetic
%
%
%
%
%
%
interactions, whatever the lattice type.

Equating $J(\vec{q},l=0)$ in the neighborhood of its maximum with
$V_c(\epsilon)$, we get the edges of the band of $q$ values
$|q|<q_c,
q_c=\frac{z}{\sqrt{2V_c(\epsilon)}}(V-\frac{V_c(\epsilon)}{z})^{\frac{1}{2}}$,
for which the $J(\vec{q})$'s are relevant. This defines a
diverging correlation length $\xi\simeq q_c^{-1}\simeq
(V-\frac{V_c(\epsilon)}{z})^{-\frac{1}{2}}$ with a mean field like
exponent $\frac{1}{2}$. We have thus established, we believe for
the first time, a diverging spatial correlation length at the
floating to coupled (classical or quantum) phase transition. This
length $\xi$ could be measured experimentally using the AC
Josephson effect. Specifically, we predict that for a 1D chain of
$N$ junctions, with opposite ends held at a small temporally
constant voltage difference $\Delta V$, an oscillating Josephson
current $I_{J}(t)$ will occur, whose frequency $\Omega$ is
{\textit {independent}} of $N$ for $N << \xi$, while being
inversely proportional to $N$ for $N >> \xi$. The crossover
between these two behaviors determines $\xi$. The linear form of
$V_c(\epsilon)=\frac{\epsilon}{A}$, where
$\epsilon=\alpha_c-\alpha$, determines the RG flows and the
perturbative phase diagram in the $V-\alpha$ plane as shown in
Fig.~\ref{Fig1}. Finally, linearizing Eq.~\ref{Jqrg} around the
critical fixed point, we get the correlation time exponent
$\nu_{\tau}$, $\xi_{\tau}\simeq |V-V_c(\epsilon)|^{-\nu_{\tau}}$,
with the continuously varying critical exponent
$\nu_{\tau}=\frac{1}{\epsilon}=\frac{1}{\alpha_c-\alpha}$. This
defines a dynamic exponent $z$, defined by $\xi_{\tau}\sim\xi^z$,
to be $z=2\nu_{\tau}=\frac{2}{\alpha_c-\alpha}$.

To reiterate the differences of the present calculation from the
case of bond dissipation \cite{Sumanta1}, we note that although
the linear order RG equation, Eq.~\ref{Vrg}, is identical in both
models, the second order equations, Eqs.~(\ref{Jrg}, \ref{Jqrg}),
are different. Because of the constancy of the number $A$, we are
able here to solve the set of infinitely many coupled non-linear
equations {\it exactly} by going to the Fourier space. This allows
us to compute important critical exponents such as $\nu$ and
$\nu_{\tau}$. We are also able here to demonstrate the divergence
of the spatial correlation length, to our knowledge for the first
time, in an FC transition. Note that, in the corresponding problem
of the coupled XY planes \cite{Toner}, or the case of bond
dissipation \cite{Sumanta1}, this was an interesting open problem
because of technical difficulties.

In systems where the `shunt to the ground' is realized, either by
a ground plane, or intrinsically in granular systems
\cite{Wagenblast, Polak}, the FC transition is tunable by varying
the shunt resistance (which varies $\alpha$ in our model).
By measuring the value of $\alpha$ at which the transition happens
for a variety of different $V$'s, it should be possible to test
our prediction for $\alpha_c$,
and that the phase boundary %
comes into the $\alpha$-axis at $\alpha_c$ with a finite, non-zero
slope. A similar phase diagram has been  obtained experimentally
in a dissipative array in Ref.~\cite{Miyazaki}, but in that
experiment a bond dissipation model was more applicable. In
systems where the present model is applicable, the non-universal
exponent $\nu_{\tau}$ can be directly measured by measuring a
characteristic temperature, $T_{\rm ch} \propto
|V-V_c(\epsilon)|^{\nu_{\tau}}$, in either phase close to the
phase boundary. In the superconducting side $T_{\rm ch}$ can be
the superconducting $T_{c}$, while in the resistive side, it can
be the temperature where the resistance shows a minimum
\cite{Fisherdiss}. On the resistive, metallic side of the
transition, the floating character can be established by measuring
the current-voltage characteristics which are expected to be a
non-uniform power law depending on $\alpha$ \cite{Kane}. Finally,
although
the correct model for capacitances of a real  experimental system
may also include longer ranged capacitors, in contrast to the
on-site capacitance $C$ used here, we can show that this does not
change any of our universal results.

We thank Sudip Chakravarty for useful discussions; JT also thanks
the Aspen Center for Physics for their hospitality while a portion
of this work was being completed. This work was supported by the
NSF under grant Nos. DMR-01-32555 and DMR-01-32726, and by
ARO-ARDA under grant No. W911NF-04-1-0236.











\end{document}